\pdfoutput=1

\documentclass[11pt,twoside,a4paper,cmspaper,final,collab]{cms-tdr}
\def\svnVersion{428928P}\def\cmsCernNoTag{CERN-EP-2016-274}\def\cmsCernDate{\today}\def\cmsMessage{Published in Physics Letters B as \href{http://dx.doi.org/10.1016/j.physletb.2017.06.053}{\doi{10.1016/j.physletb.2017.06.053.}}}

\begin{document}\cmsNoteHeader{HIN-15-012}

\hyphenation{had-ron-i-za-tion}
\hyphenation{cal-or-i-me-ter}
\hyphenation{de-vices}
\RCS$Revision: 428913 $
\RCS$HeadURL: svn+ssh://svn.cern.ch/reps/tdr2/papers/HIN-15-012/trunk/HIN-15-012.tex $
\RCS$Id: HIN-15-012.tex 428913 2017-10-11 13:00:41Z kjung $
\newlength\cmsFigWidth
\ifthenelse{\boolean{cms@external}}{\setlength\cmsFigWidth{0.98\columnwidth}}{\setlength\cmsFigWidth{0.85\textwidth}}
\ifthenelse{\boolean{cms@external}}{\providecommand{\cmsLeft}{top\xspace}}{\providecommand{\cmsLeft}{left\xspace}}
\ifthenelse{\boolean{cms@external}}{\providecommand{\cmsRight}{bottom\xspace}}{\providecommand{\cmsRight}{right\xspace}}
\providecommand{\sqrtsNN}{\ensuremath{\sqrt{s_{\mathrm{NN}}}}\xspace}
\newcommand{\etaCM}{\ensuremath{\eta_{\mathrm{CM}}}\xspace}
\newcommand{\etalab}{\ensuremath{\eta_{\text{lab}}}\xspace}
\newcommand{\RpA}{\ensuremath{R_{\Pp\mathrm{A}}}\xspace}

\cmsNoteHeader{HIN-15-012}
\title{Measurements of the charm jet cross section and nuclear modification factor in pPb collisions at $\sqrtsNN$ = 5.02\TeV}

\date{\today}

\abstract{
	The CMS Collaboration presents the first measurement of the differential cross section of jets from charm quarks produced in proton-lead (pPb) collisions at a
	nucleon-nucleon center-of-mass energy of $\sqrtsNN = 5.02\TeV$,
	as well as results from charm quark jets in proton-proton (pp) collisions at $\sqrt{s} = 2.76$ and 5.02\TeV. By comparing the yields
		   of the pPb and pp collision systems at the same energy,
		   a nuclear modification factor for charm jets from 55 to 400\GeVc in pPb collisions at $\sqrtsNN = 5.02\TeV$ of $R_{\Pp\mathrm{A}} = 0.92\pm0.07\stat\pm0.11\syst$ is obtained.  This is consistent
		   with an absence of final-state energy loss for charm quarks in pPb collisions. In addition, the fraction of jets coming from charm quarks
		   is found to be consistent with that predicted by \PYTHIA 6 for pp collisions at $\sqrt{s} = 2.76$ and 5.02\TeV, and is independent of the jet transverse momentum from 55 to 400\GeVc.
}

\hypersetup{%
pdfauthor={CMS Collaboration},%
pdftitle={Measurements of the charm jet cross section and nuclear modification factor in pPb collisions at sqrt(s[NN]) = 5.02 TeV},%
pdfsubject={CMS},%
pdfkeywords={CMS, heavy ions, charm-tagging, heavy-flavor}}

\maketitle

\section{Introduction}
The creation of a new state of matter, known as quark-gluon plasma (QGP), has been predicted by lattice calculations for states of matter with extremely high energy densities~\cite{Karsch:2001vs}.  Collisions of heavy nuclei studied at both the BNL RHIC and CERN LHC facilities have been observed to create energy densities larger than that required for QGP creation~\cite{Adams:2005dq,LHCQGPSummary,HIN-11-003,phenixdEdEta}.
The QGP is a state of matter which is characterized by an effective deconfinement of the quark and gluon color degrees of freedom. Hard-scattered partons are expected to lose energy via elastic and inelastic interactions as they traverse the QGP~\cite{Bjorken:1982tu}. This is commonly thought to be the mechanism responsible for the observed suppression of high transverse momentum (\pt) hadrons and jets, or ``jet quenching'', in nuclear collisions~\cite{Adcox:2004mh, Adams:2005dq, Back:2004je, Arsene:2004fa, Chatrchyan:2011sx, Aad:2010bu, ALICEJetQ, ALICEJetQ2}.

Jet quenching is expected to depend on the flavor of the fragmenting parton~\cite{vanHees:2005aa,teaney:2005aa}, primarily due to two effects:
first, heavy quarks may suffer mass-dependent effects further separating their energy loss measurements from those of inclusive jets.  For example, it is expected that the radiative and collisional energy loss mechanisms should have different strengths for heavy quark and light quark jets~\cite{vitev:2013aa, dokshitzer:2001aa}. Therefore, heavy quarks can provide new information on the relative jet quenching power of these various energy loss mechanisms.
Second, a pure heavy flavored jet sample does not generally contain jets seeded by high-\pt gluons, contrary to a measurement of inclusive jets, which contains a sizable gluon-jet component as predicted by \PYTHIA~\cite{Sjostrand:2006za} simulations. Under the assumption that gluon radiation is the dominant mechanism for energy loss, gluon jets are expected to quench more strongly than quark jets, owing to the larger color factor for gluon emission from gluons than from quarks~\cite{denterria:2013}.  By identifying charm and bottom jets (c and b jets), measurements can be performed on a jet sample with an enhanced fraction of quark jets.

The energy loss discrimination power of both effects is mitigated somewhat due to the presence of gluon splitting, which is a next-to-leading order heavy quark production mechanism where a high-energy gluon can split into a quark pair. At high-\pt, the heavy flavored quark production fraction from gluon splitting is expected to be roughly 50\%~\cite{banfi:2007}, but as the gluon virtuality is also quite large, it may be the case that the quarks from gluon splitting still experience the majority of the QGP medium evolution.

The CMS Collaboration has also previously observed QGP effects on heavy-flavored objects through measurements of fully-reconstructed mesons~\cite{BMesonCMS}. While meson measurements are able to access the low-\pt regime in a more effective way than jets, the measurements are less direct as a result of the fragmentation process. In other words, the connection to the b or c quark energy loss is smeared by its combination with a light quark to create the reconstructed object, whereas jets aim to capture the entire energy of the fragmenting quark.

Previous measurements of jets in proton-lead (pPb) collisions have not observed significant jet quenching effects~\cite{HIN-13-001,HIN-14-001,AtlasRpA,ALICEjetRpA}, suggesting that measurements from pPb collisions can place limits on the extent of ``cold nuclear matter" effects on jet production~\cite{salgado:2011}.  One such initial-state effect is due to the nuclear parton distribution functions (nPDFs).  These nPDFs are expected to enhance the charm quark yields by roughly 10--15\%, as the kinematic selections used in this analysis correspond to the ``antishadowing" region of the Bjorken-x distribution~\cite{eskola:2016}.
While the modification factors \RpA~for both b jets~\cite{HIN-14-007} and inclusive jets~\cite{HIN-14-001} at a nucleon-nucleon center of mass energy of $\sqrtsNN = 5.02\TeV$ have been measured by CMS, these measurements used a \PYTHIA simulation and an interpolated pp reference as baselines, respectively, as at the time of publication, no 5.02\TeV proton-proton (pp) data was available. This analysis presents the first measurement of an inclusive charm jet cross section in pPb collisions at $\sqrtsNN = 5.02\TeV$, including comparisons to the cross sections in pp collisions at both $\sqrt{s} = 2.76$ and 5.02\TeV.

\section{Detection, Reconstruction, and Simulation}

\subsection{Detection}
The CMS detector has excellent capabilities to perform displaced jet identification (b and c tagging) as demonstrated in Ref.~\cite{Chatrchyan:2012jua}. The central feature of the CMS apparatus is a superconducting solenoid of 6\unit{m} internal diameter, providing a magnetic field of 3.8\unit{T}. Within the solenoid volume are a silicon pixel and strip tracker, a lead tungstate crystal electromagnetic calorimeter, and a brass and scintillator hadron calorimeter (HCAL), each composed of a barrel and two endcap sections. Extensive forward calorimetry complements the coverage provided by the barrel and endcap detectors.
The tracker has a pseudorapidity coverage of $\abs{\etalab} < 2.4$, while the calorimetry covers $\abs{\etalab} < 3$. Muons are measured in gas-ionization detectors embedded in the steel flux-return yoke outside the solenoid. A more detailed description of the CMS detector, together with a definition of the coordinate system used and the relevant kinematic variables, can be found in Ref.~\cite{Chatrchyan:2008zzk}.

Event selections are identical to previous pPb analyses \cite{HIN-12-017,HIN-14-001,HIN-14-007} and include the requirement of a primary vertex within 15\unit{cm} of the nominal interaction point in the beam direction and the removal of events consisting primarily of HCAL noise. Beam-related background is suppressed by rejecting events in which less than 25\% of all reconstructed tracks are of good quality.

\subsection{Reconstruction}
Jets are reconstructed offline using the particle-flow algorithm~\cite{CMS-PAS-PFT-10-001}, which identifies each individual jet constituent as one of a number of different particle types, including photons, electrons, muons, charged hadrons, and neutral hadrons.  This is done using an optimized combination of information from the various elements of the CMS detector~\cite{particle_flow}. These particle-flow candidates do not have explicit kinematic selections, though charged tracks are limited to $\pt > 400$\MeV. Jets are clustered by the anti-\kt algorithm~\cite{bib_antikt} with a radius of 0.3.
Jet energy corrections are derived from simulation and using measurements of energy balance in dijet and photon+jet events.  Finally, an iterative underlying event removal procedure is applied to jets in pPb events~\cite{kodolova:2007}.  Jet momentum is found from
simulation to be within 2\% of the true jet momentum over the whole \pt spectrum and
detector acceptance after the jet energy corrections are applied for both pp and pPb collisions. This residual nonclosure is primarily due to differing jet energy resolution between quark and gluon jets.

Three different data sets collected by the CMS experiment are used, corresponding to integrated luminosities of 35\nbinv of pPb collisions at $\sqrtsNN = 5.02$\TeV and 4.8\pbinv of pp collisions at $\sqrt{s} = 2.76\TeV$ taken during the 2013 heavy ion run period at the LHC, as well as 27.9\pbinv of pp collisions at $\sqrt{s} = 5.02$\TeV collected during the 2015 heavy ion run period. During the pPb run, the proton and lead beam energies per nucleon were different, which led to a center-of-mass pseudorapidity ($\eta$) shift of 0.465 units with respect to the laboratory frame. After an integrated luminosity of 20.9\nbinv was collected, the directions of the proton and lead beams were reversed.  In this analysis, the beam parameters are redefined such that the proton beam is always traveling in the positive $\eta$ direction. Therefore, the laboratory and the center-of-mass pseudorapidities are related as $\etalab = \etaCM + 0.465$.

As jet energy corrections are only reliable for $\pt >20$\GeVc, single jets are required to have a raw online \pt above that cutoff and a fully-corrected $\pt>35$\GeVc.  In order to mitigate effects from the limited CMS inner tracker $\eta$ acceptance of $\abs{\etalab} < 2.4$ and the boost between the lab and center-of-mass reference frames, jets in pPb collisions are required to be reconstructed within $\abs{\etaCM} < 1.5$, while jets in pp collisions can be found within $\abs{\etaCM} <2.0$. When direct comparisons of quantities in pp and pPb collisions are shown, jets from both systems use a pseudorapidity selection of $\abs{\etaCM} < 1.5$.

Events are selected online by one or more jet triggers with varying energy thresholds.  In the 2.76\TeV pp and 5.02\TeV pPb analysis, five single-jet triggers with \pt thresholds of 20, 40, 60, 80, and 100\GeVc are combined in order to maximize the number of accepted events over a wide range of jet \pt.  As some lower \pt triggers are prescaled, meaning that a fraction of the triggered events are randomly rejected to constrain data throughput, a simple OR of all triggers will bias the jet \pt spectrum toward the larger threshold triggers and will also have significant event duplication. Instead, a trigger combination procedure based on the trigger prescale factors is used. This trigger combination is also used in the analysis of b jets in pPb \cite{HIN-14-007} and is briefly described here.  The jet with the largest online raw \pt, i.e. the \pt used by the triggers before jet energy corrections, is used to classify each event.  Based on this online raw jet \pt, it is possible to deduce which triggers have been satisfied, irrespective of whether a trigger is prescaled.  If the highest fired trigger conditions are satisfied, the event is kept and weighted by the corresponding trigger prescale factor, else the event is discarded. After this combination, the jet finding efficiency of the full sample is $>$99.9\% for jets above 35\GeVc, and the total event selection efficiency is around 97\%.

For the 5.02\TeV pp data, the trigger menu was slightly altered in preparation for the higher instantaneous luminosity achieved in the 2015 run period, so only four triggers are combined with \pt thresholds of 40, 60, 80, and 100\GeVc.  As a result of jet energy smearing effects from reconstruction and resolution unfolding, the absence of a 20\GeVc trigger effectively places a 55\GeVc~lower bound on the leading jet \pt for the 5.02\TeV pp data, rather than the roughly 40\GeVc~bound at 2.76\TeV.

\subsection{Simulation}
This analysis relies on simulations of pp collisions at 2.76 and 5.02\TeV, as well as simulations of pPb collisions at 5.02\TeV. Monte Carlo (MC) simulations of inclusive quantum chromodynamic (QCD) hard-scattering events are generated using \PYTHIA 6.424~\cite{Sjostrand:2006za}, tune Z2~\cite{pythiaz2}.
These events are generated imposing thresholds on the transverse momentum of the hard scattering subprocess ($\hat{p}_{\mathrm{T}}$) in order to force production of jets with high \pt.  In order to properly build templates, unfold the jet resolution, and calculate the tagging efficiency in the proton-nucleus environment, minimum-bias pPb events are produced using the \HIJING1.383 event generator~\cite{bib_hijing} at $\sqrtsNN = 5.02$\TeV.  Simulated events from \PYTHIA 6 are produced at 5.02\TeV in conjunction with a pPb background event.  In this way, each simulated pPb event contains at least one jet produced by a hard scattering subprocess while still accurately representing the jet resolution and energy scale in a pPb environment. To account for possible differences in reconstruction performance between the two boost directions, MC samples were obtained for both directions of the proton beam.  For pp collisions, \etalab is identical to \etaCM. Jets generated by the \HIJING simulation of the underlying pPb events are rejected in the analysis since these jets can be quenched~\cite{bib_hijing}, possibly resulting in a modified fragmentation pattern which would bias the jet energy corrections. Within the kinematic selections of the analysis, the jets from \HIJING account for less than 1\% of the total jet fraction.

\section{Charm quark tagging}

In Monte Carlo studies, a charm jet is defined as any jet containing a prompt charm quark within the jet cone and ignoring jets which contain a b$\to$c cascade decay.  Identification of such jets is achieved by tagging vertices consistent with decays of hadrons containing a charm quark. Even though the maximum displacement of such charmed hadron decays is only on the order of 100\mum for the kinematic selections of this analysis, the presence of a silicon tracker very close to the interaction point at CMS allows for the discrimination of secondary vertices with such small displacement values.  For proton-proton collisions, individual track vertexing uncertainties in the beam direction are on the order of 100\mum at 1\GeVc and 40\mum at 10\GeVc, while the uncertainties in the transverse direction are on the order of 70\mum at 1\GeVc and 20\mum at 10\GeVc~\cite{CMStrking}.

\begin{figure*}[htbp]
\centering
\includegraphics[width=\textwidth]{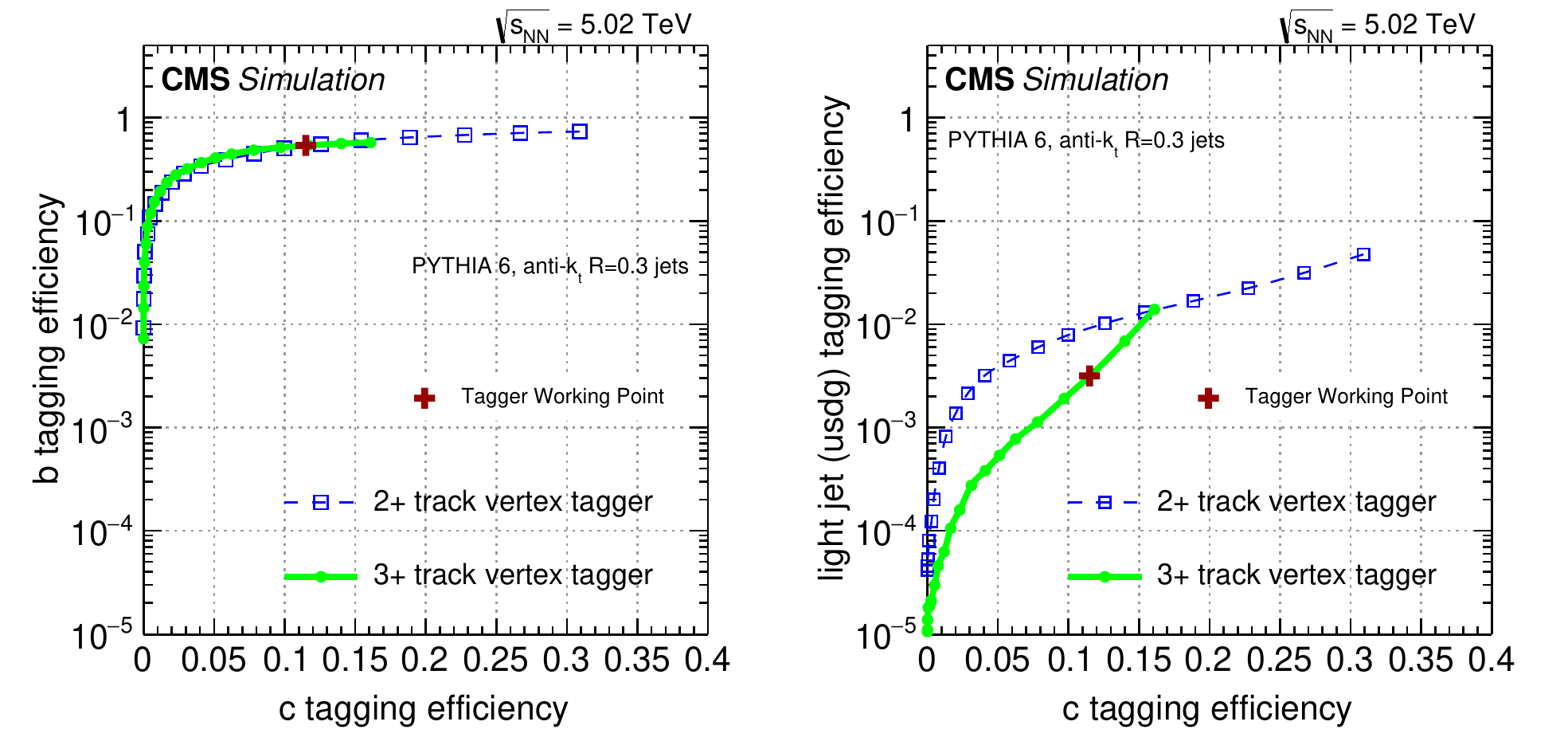}
\caption[Performance of the Simple Secondary Vertex tagger]{\label{fig:SSVeffComp} Efficiency of tagging b jets (left) and light parton jets (right) for the high-purity (3+ track), and high-efficiency (2+ track) versions of the simple secondary vertex (SSV) tagger as a function of c jet tagging efficiency.  The charm-to-bottom discrimination power is virtually unchanged between the high-efficiency and high-purity versions of the tagger, while the light parton jet mistag rate is reduced by a factor of three at the analysis working point, shown as the closed red cross on the plots. (For interpretation of the references to color in this figure, the reader is referred to the web version of this article.)
}
\end{figure*}

\begin{figure}[htbp]
\centering
\includegraphics[width=0.49\textwidth]{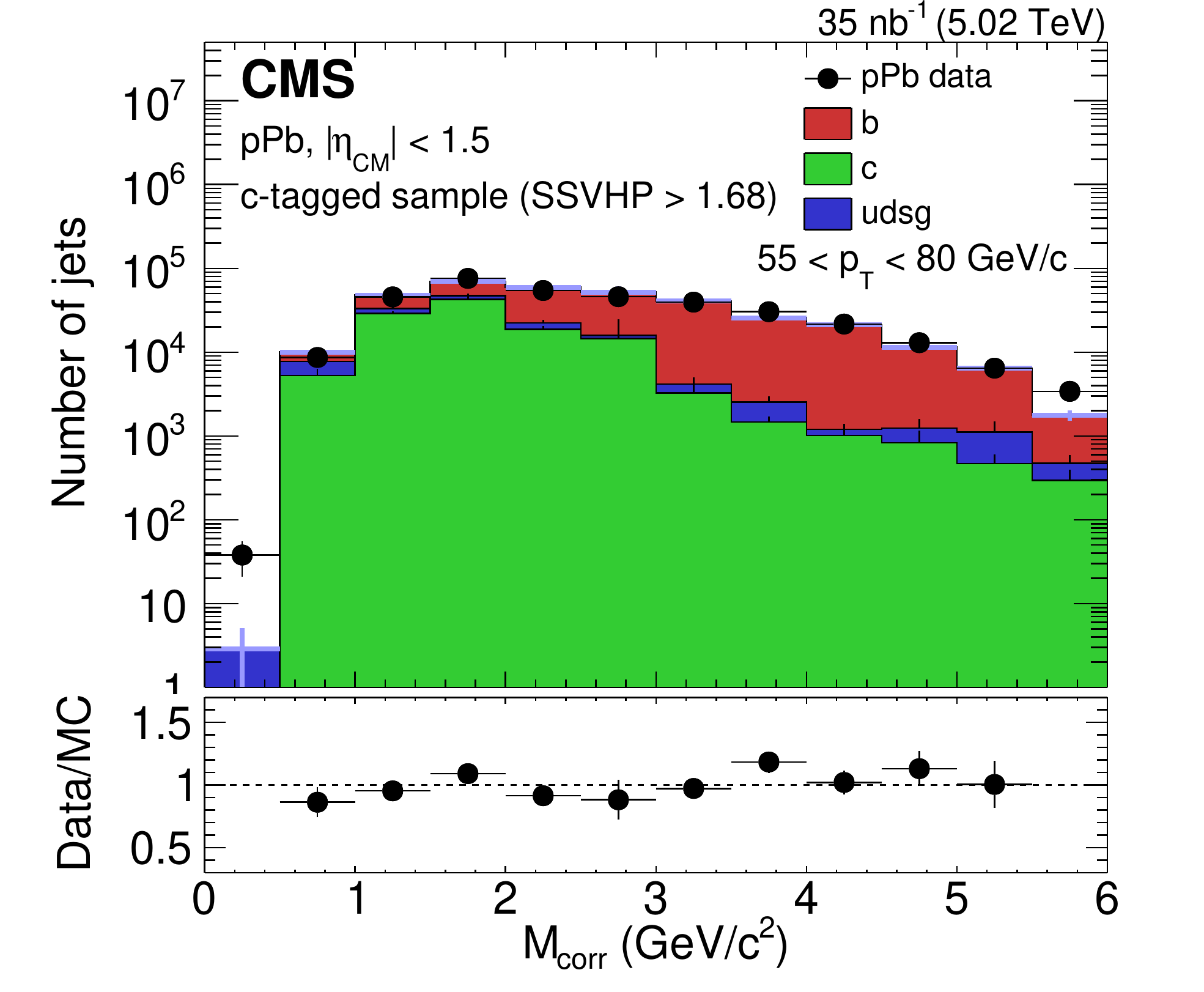}
\includegraphics[width=0.49\textwidth]{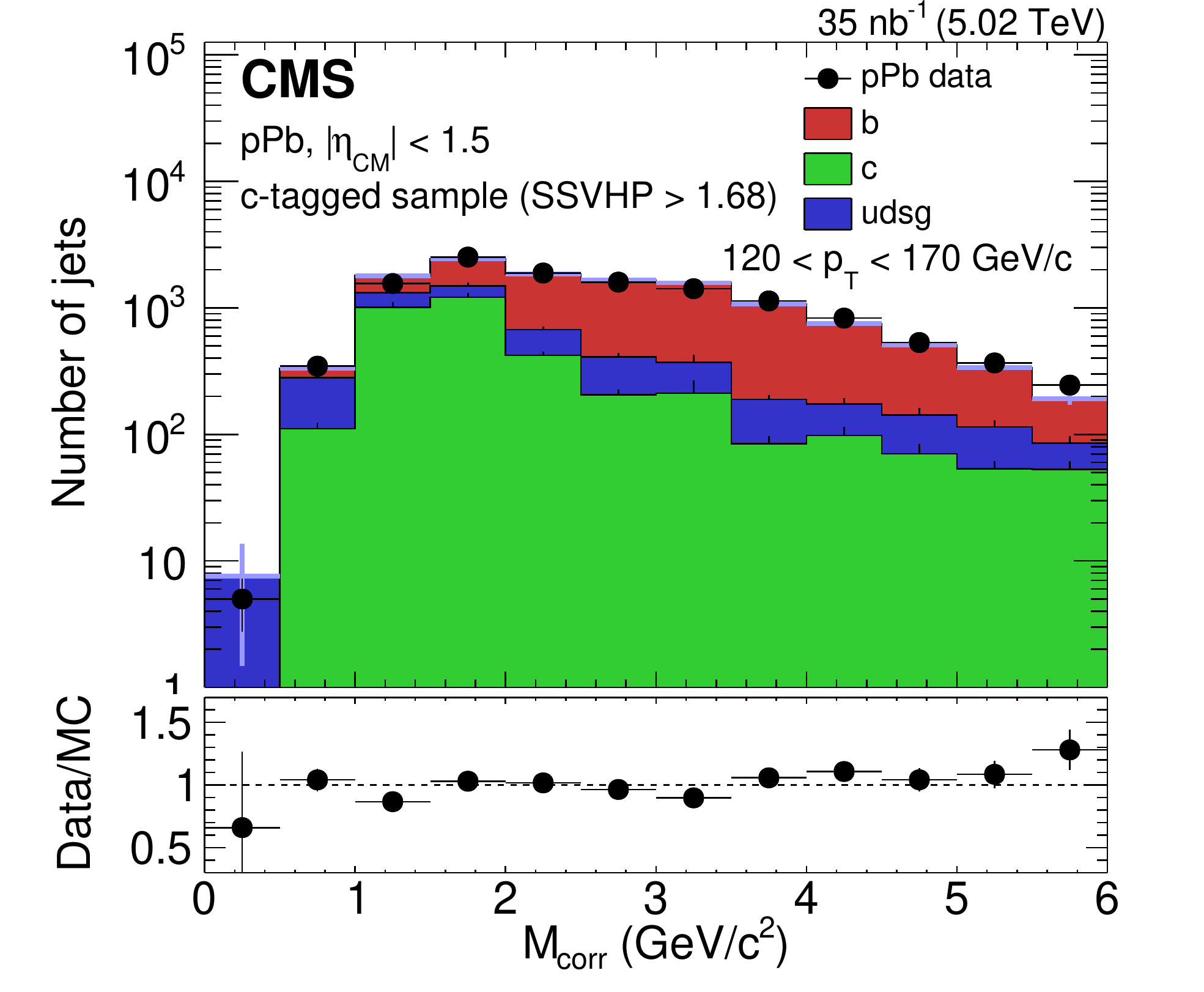}
\caption[Corrected secondary vertex mass template fits]{\label{fig:svtxmCorrFit}
Corrected secondary vertex mass distributions from {\PYTHIA{+}\HIJING} for c jets (green), light parton jets (blue), and b jets (red) in the jet \pt range 55--80\GeVc (\cmsLeft) and 120--170\GeVc (\cmsRight).  Relative normalizations of these three distributions are fit to a distribution from pPb collision data (black).  Statistical uncertainties are shown in black for data and for individual simulated flavor components and are shown in blue for the sum of the simulated distributions. The bottom panels of both plots show the ratio of data to simulation. (For interpretation of the references to color in this figure, the reader is referred to the web version of this article.)
}
\end{figure}

\begin{figure}[htbp]
\centering
\includegraphics[width=0.49\textwidth]{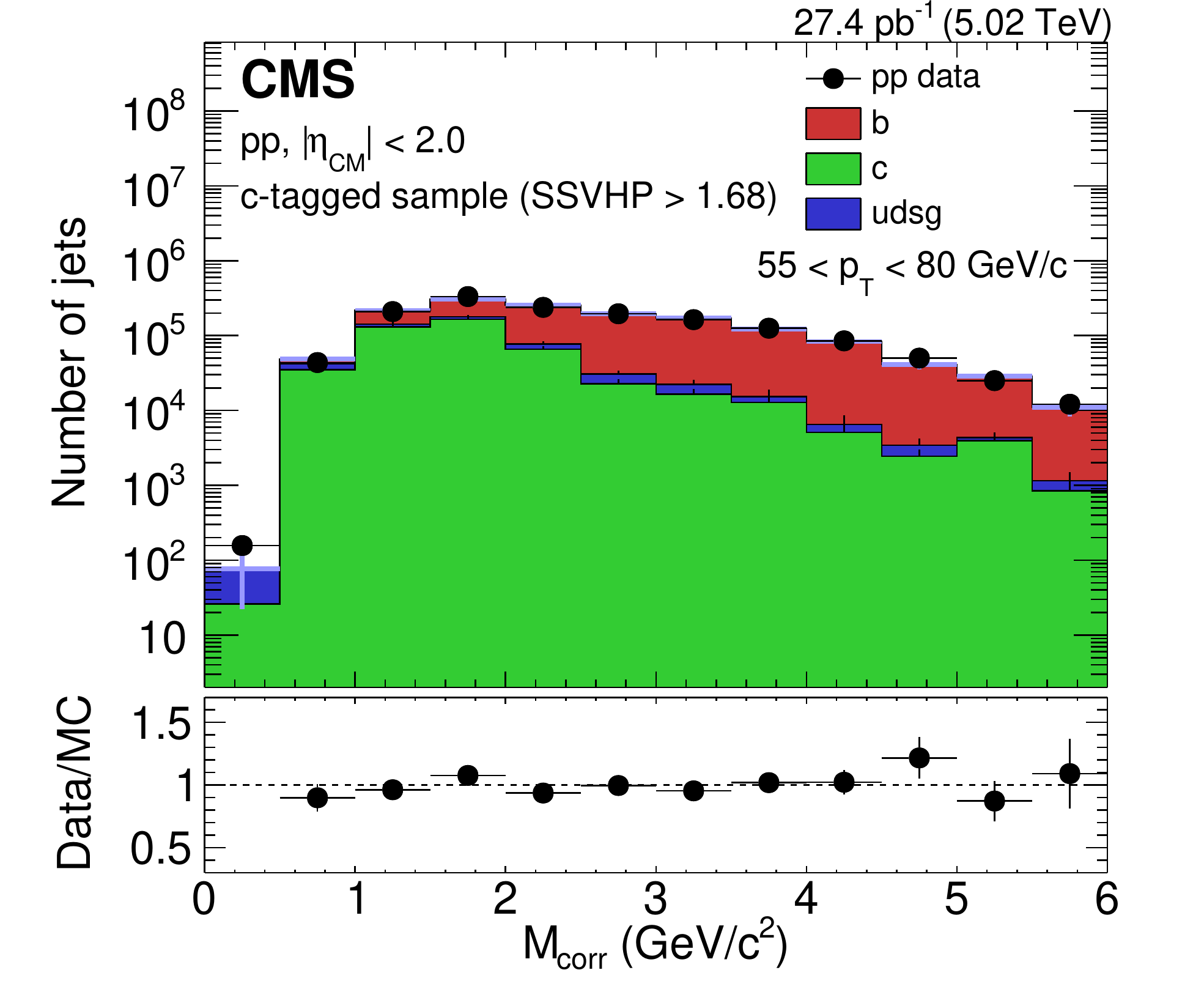}
\includegraphics[width=0.49\textwidth]{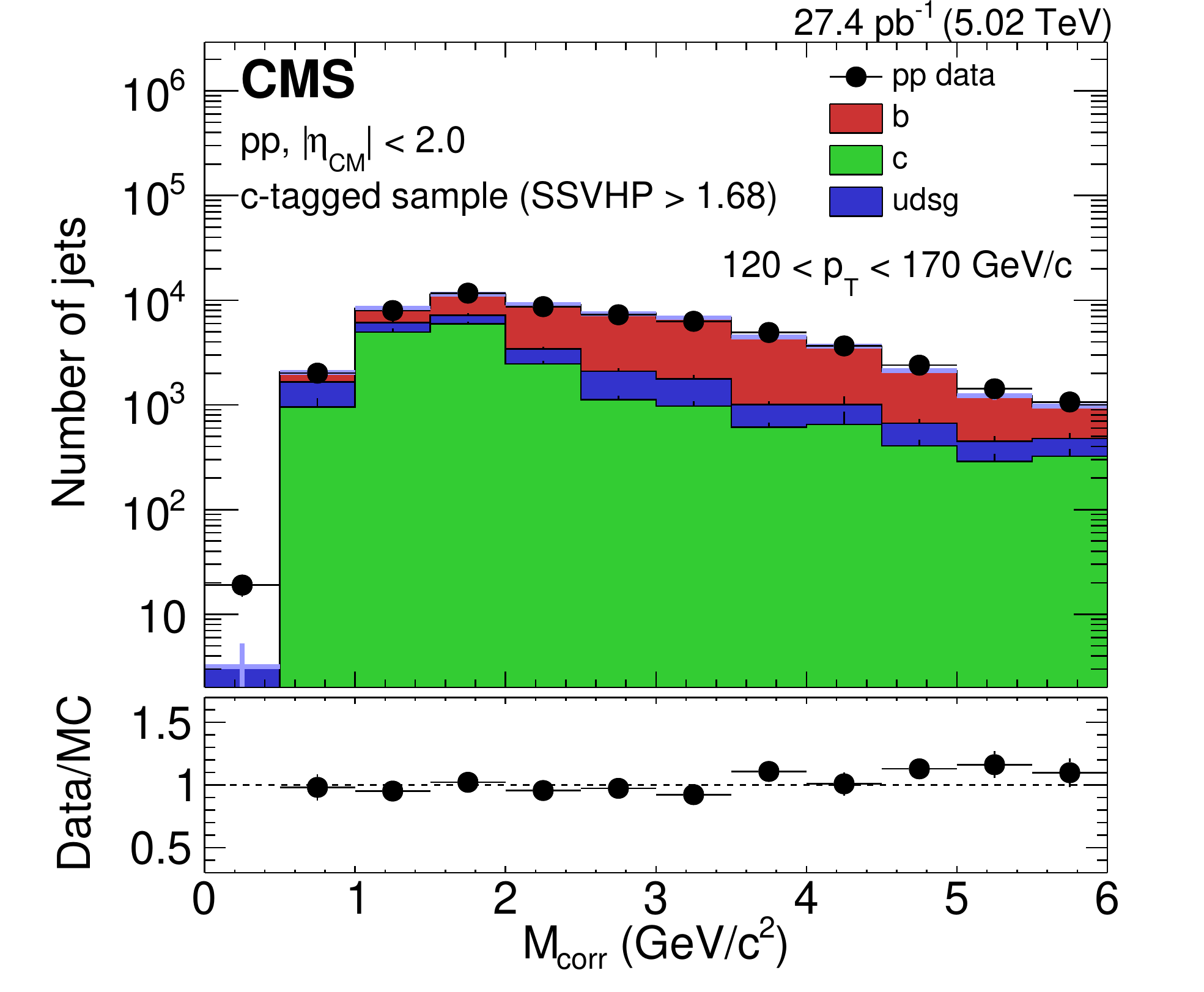}
\caption[Corrected secondary vertex mass template fits]{\label{fig:svtxmCorrFit_5TeVpp}
Corrected secondary vertex mass from a \PYTHIA 6, tune Z2 simulation for c jets (green), light parton jets (blue) and b jets (red) in the jet \pt range 55--80\GeVc (\cmsLeft) and 120--170\GeVc (\cmsRight).  Relative normalizations of these three distributions are fit to a distribution from pp collision data (black).  Statistical uncertainties are shown in black for data and for individual simulated flavor components and are shown in blue for the sum of the simulated distributions. The bottom panels of both plots show the ratio of data to simulation. (For interpretation of the references to color in this figure, the reader is referred to the web version of this article.)
}
\end{figure}

\begin{figure}[htbp]
\centering
\includegraphics[width=0.49\textwidth]{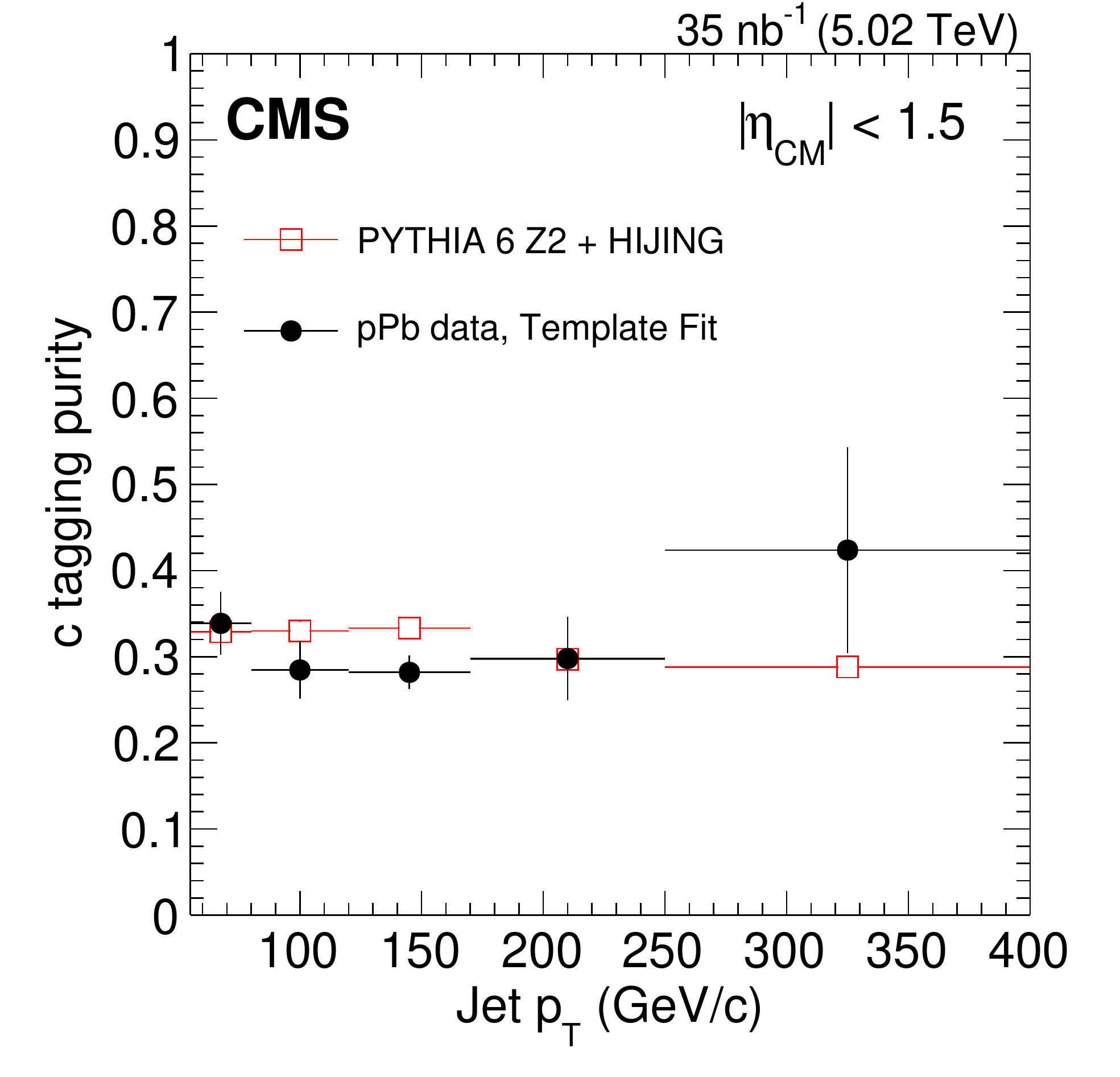}
\includegraphics[width=0.49\textwidth]{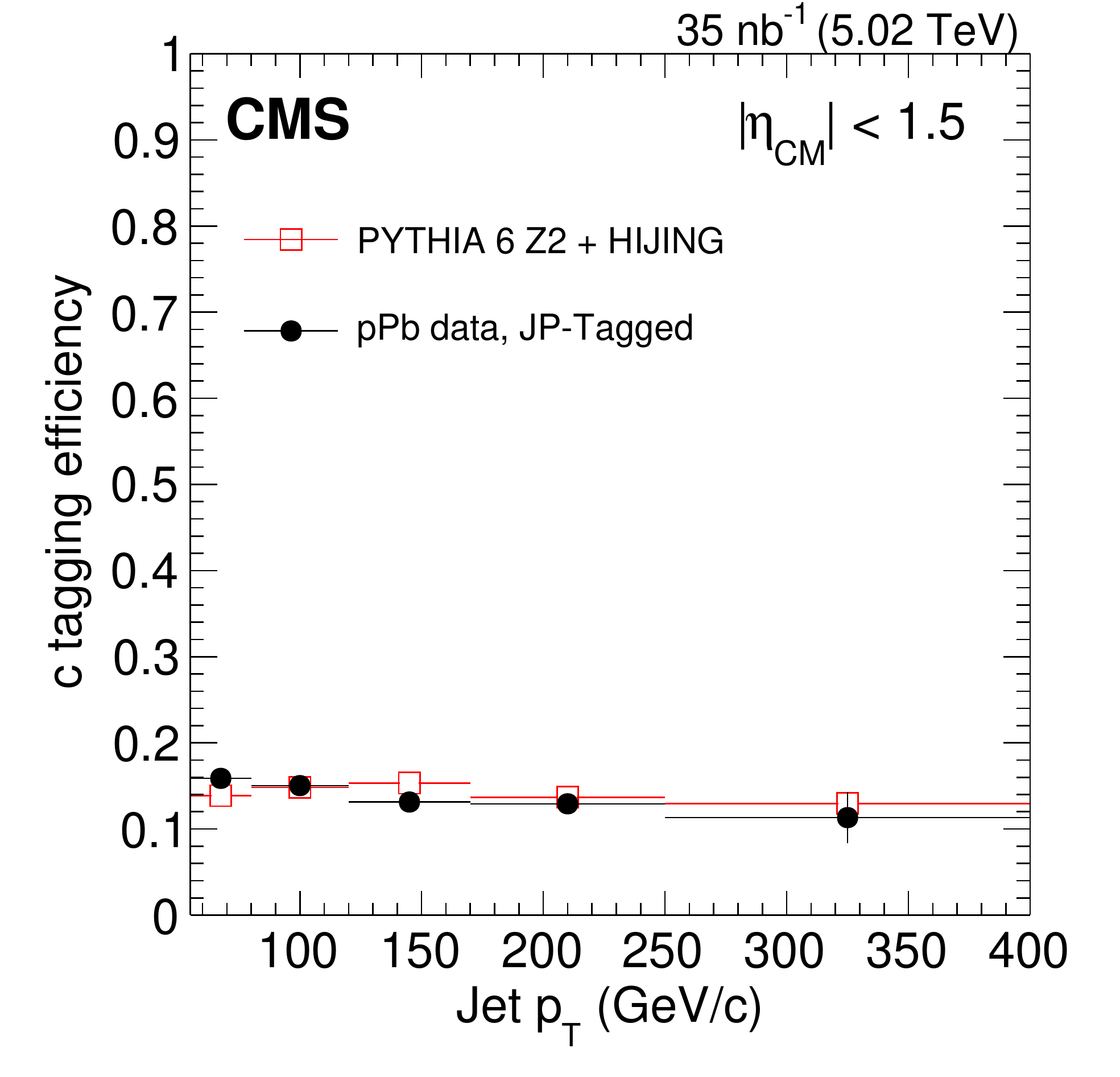}
\caption[Efficiency and purity of charm tagging in pPb]{\label{fig:PurandEff}
Tagging purity (\cmsLeft) and efficiency (\cmsRight) for the working point selection of $\mathrm{SSVHP} > 1.68$ in pPb collisions at 5.02\TeV for simulation (open red squares) and data (closed black points). (For interpretation of the references to color in this figure, the reader is referred to the web version of this article.)
}
\end{figure}

\begin{figure}[htbp]
\centering
\includegraphics[width=0.49\textwidth]{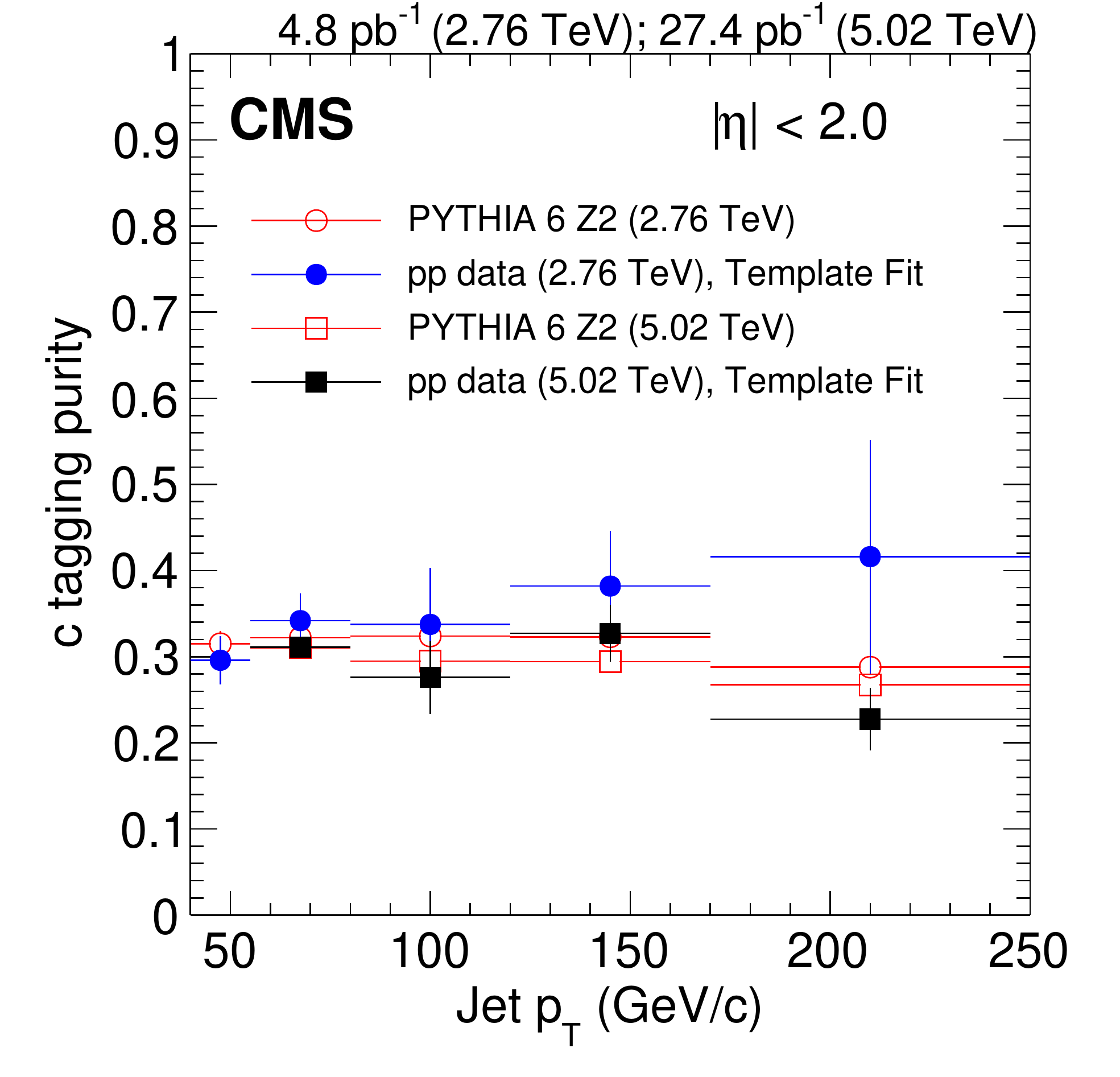}
\includegraphics[width=0.49\textwidth]{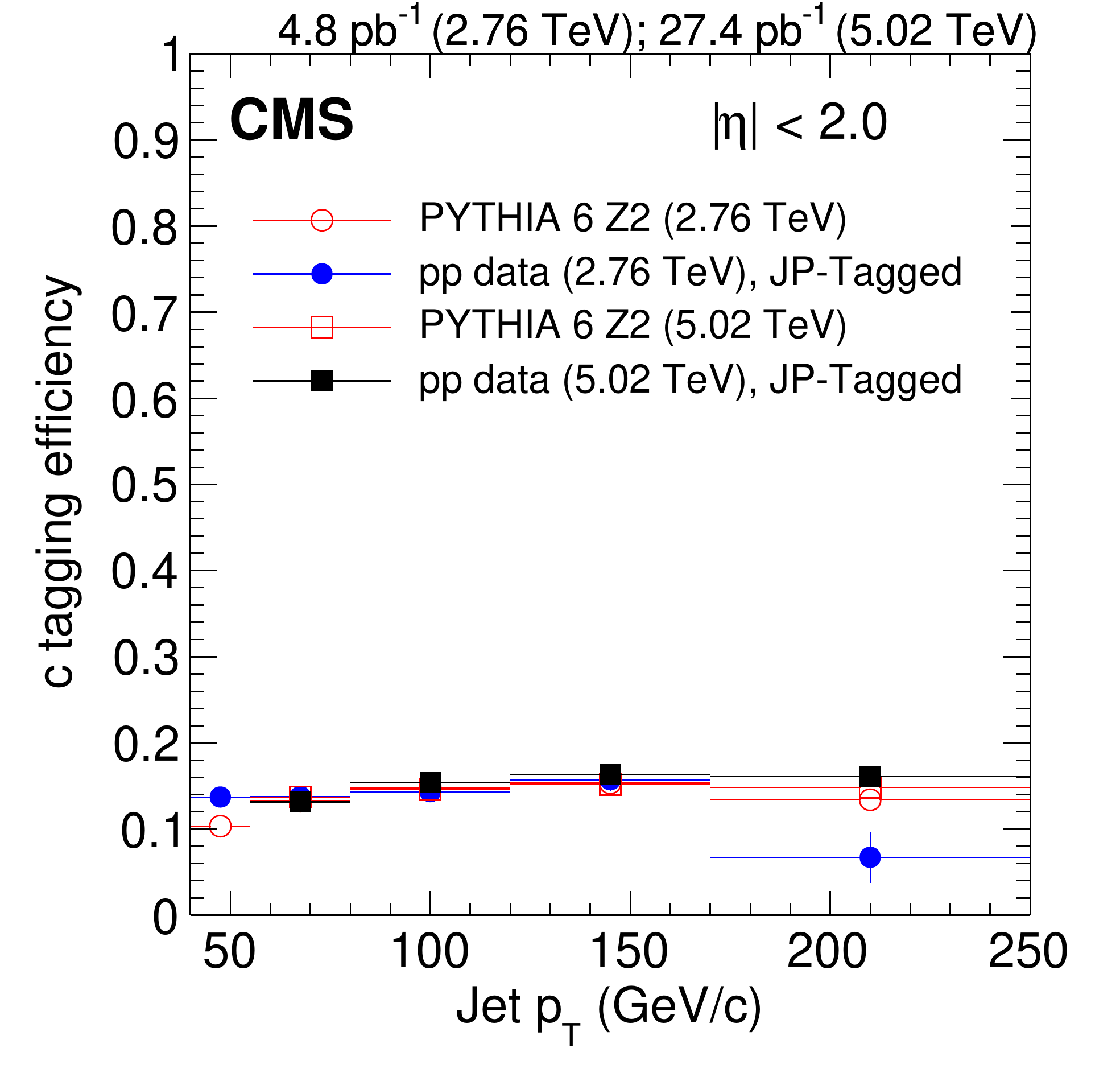}
\caption[Efficiency and purity of charm tagging in pp]{\label{fig:PurandEff_5TeVpp}
The tagging purity (\cmsLeft) and efficiency (\cmsRight) for the working point selection of $\mathrm{SSVHP} > 1.68$ in pp collisions at 5.02\TeV (square markers) and at 2.76\TeV (circular markers). Purity curves from simulation (open red markers) and data (closed markers) are shown, obtained by fitting templates to the data. The right plot shows efficiency curves from simulation (open red markers) and the cross-check based on JP tagging. (For interpretation of the references to color in this figure, the reader is referred to the web version of this article.)
}
\end{figure}

This c jet analysis closely follows previous CMS analysis strategies for heavy-flavor jet identification, or tagging, specifically the measurements of b quark jets in heavy ion environments in CMS, both in lead-lead collisions~\cite{HIN-12-003} and pPb collisions~\cite{HIN-14-007}.
This analysis strategy uses two different taggers to identify c jets.  While both taggers assign a numerical discriminator quantifying how ``charm like" each jet is, each tagger uses a slightly different identification strategy.  The first tagger is known as the simple secondary vertex (SSV) tagger~\cite{Chatrchyan:2012jua} and uses reconstructed displaced vertices. The version of the SSV tagger used in this analysis is the ``high-purity" (SSVHP) one, which requires the presence of a secondary vertex in the jet cone with at least three associated tracks, each with track $\pt>1$\GeVc. All versions require that all secondary vertices share fewer than 20\% of tracks with any other vertex.
The inclusion of the third associated vertex track in the high-purity version of the tagger allows for the selection of a tagging working point that reduces the misidentification rate of light jets by a factor of three, while still keeping a large majority of c jets, as shown in Fig.~\ref{fig:SSVeffComp}.  With a reduced light jet contamination, c jets begin to dominate small regions of kinematic phase space, which this analysis exploits to extract relative flavor contributions of light, c, and b jets to the total jet sample.

The second tagger used in this analysis is known as the jet probability (JP) tagger~\cite{Chatrchyan:2012jua}, and is used to cross-check the tagging efficiency predicted by simulation using control samples in data.  This tagger uses a numerical discriminator based on the presence of single tracks that are significantly displaced from the primary vertex, and is therefore largely uncorrelated with secondary vertex reconstruction performance.
The efficiency of a particular tagger (e.g. SSVHP) can be calculated with the JP tagger:

\begin{equation}
\label{eq:JPeff}
\epsilon_{\text{tag}} = \frac{C_{\PQc}f_{\PQc}^{\text{tagged}}N_{\rm{jets}}^{\text{tagged}}}{f_{\PQc}^{\rm{pretag}}N_{\rm{jets}}^{\rm{pretag}}},
\end{equation}

where $f_{\PQc}^{\text{tagged}}$ is the purity of the sample from a JP discriminator template fit after applying the SSVHP discriminator selection, and $f_{\PQc}^{\rm{pretag}}$ is the same but before this selection, $N_{\rm{jets}}^{\rm{pretag}}$ and $N_{\rm{jets}}^{\text{tagged}}$ denote the number of jets before and after tagging, respectively, and $C_c$ denotes the fraction of jets that can be identified by the JP tagger (generally very close to one).

The tagging efficiency is calculated both from simulation and using distributions of the JP tagger~\cite{Chatrchyan:2012jua} both before and after imposing the SSVHP tagging requirement.  A unique advantage of using the JP tagger for calculating tagging efficiency via Eq. (\ref{eq:JPeff}) is that it can be calibrated using data to correct for the effects of tracking resolution.  Tracks with negative values of impact parameter significance (i.e.~tracks with vertex displacements on the away-side of the vertex from the jet) are purely a product of resolution smearing and these can be used to compute a probability for the association of any given track to the primary vertex.  The tagger distributions are calibrated independently in data and simulation such that the distribution of negative impact parameters is flat (by construction) as a function of track displacement. Through the calibration of the JP tagger, the impact parameter significance distributions in both data and simulation are transformed from unbounded into bounded distributions, such that both can be analyzed on an equal footing. Once recalibrated, the residual difference between the tagging efficiency derived from simulation and from the JP calculation (Eq.~\ref{eq:JPeff}) is used as the systematic uncertainty estimation. 

The c jet purity calculation relies on another discriminating variable known as the corrected secondary vertex mass.  This was first developed as a tool for identifying b jets by the experiments at LEP~\cite{LEP_bquark} and SLC~\cite{SLAC_svReco} and is also used by the LHCb Collaboration \cite{LHCb_cTag}. The motivation behind this variable is to correct for any missing mass of the decay vertex due to neutral or unobserved particles.  If the momentum vector of the collection of particles associated to a vertex is not parallel to the vector pointing from the primary vertex to the secondary vertex decay point, i.e. the flight direction of the constituent particles, one can use conservation of momentum to calculate a minimum possible mass the vertex must have had.
This minimum possible mass is called the corrected secondary vertex mass, or $M_\text{corr}$, and is defined as:

\begin{equation}
\label{eq:svtxmcorr}
M_\text{corr} = \sqrt{M^{2}+(p/c)^{2}\sin^{2}\theta} + (p/c)\sin\theta,
\end{equation}

where $M$ is the invariant mass of the vertex, $p$ is the momentum of the vector sum of the reconstructed particles that form the secondary vertex, and $\theta$ is defined as the angle between that summed momentum vector and the flight direction of the vertex.  If all particles that belong to a given secondary vertex are reconstructed, the angle $\theta$ should be zero, and the secondary vertex mass needs no correction.  Otherwise, the value of $M_\text{corr}$ is used in the calculation of the vertex mass to account for the nonreconstructed momentum.

The c jet purity is found using template fits of $M_\text{corr}$, after using the SSVHP tagger.
The numerical values of the SSVHP discriminator are correlated to the significance of the secondary vertex displacement with respect to the primary vertex and are obtained using the formula: SSVHP = ln(1 + $\abs{d}/\sigma(d)$), where $d$ is the three-dimensional vertex displacement and $\sigma(d)$ is the uncertainty in the displacement measurement.  The working point used in this analysis requires SSVHP $>$ 1.68, which maximizes the estimated c jet purity from the MC samples, increasing the c jet purity from around 10\% to around 30\%. Once the working point selection is applied to the sample, distributions of corrected secondary vertex mass from light parton, c, and b jets in the \PYTHIA{+}\HIJING or \PYTHIA simulations are fit to distributions in data.
The shapes of the different flavor templates are fixed, but the relative normalizations of each flavor template are allowed to float independently.  As seen in Fig.~\ref{fig:svtxmCorrFit} for pPb collisions, and in Fig.~\ref{fig:svtxmCorrFit_5TeVpp} for pp collisions at 5.02\TeV, b jets dominate the $M_{\rm{corr}}$ distributions for vertex masses above 3 \GeVcc, while the light parton jet contribution is significantly reduced by the SSVHP tagger requirement.  Because of this light parton jet removal, the relative c~jet contribution to the sample below 3 \GeVcc is quite large, allowing for an accurate extraction of the c jet purity in the data sample.

Figure~\ref{fig:PurandEff} shows the c tagging purity and efficiency of the sample after applying the SSVHP tagger selection for 5.02\TeV pPb collisions, both in data and simulation. Figure~\ref{fig:PurandEff_5TeVpp} depicts the same for 5.02 and 2.76\TeV pp collisions, again, both in data and simulation.

Once the efficiency and purity values are found, the total number of c jets in the sample is obtained \pt bin by \pt bin using:

\begin{equation}
\label{eq:Ncjets}
N_{\text{\PQc jets}} = N^{\text{tagged}}_{\text{jets}}\frac{f_{\PQc}}{\epsilon_{\text{tag}}},
\end{equation}

where $N^{\text{tagged}}_{\text{jets}}$ is the number of jets passing the SSVHP working point selection, $f_{\PQc}$ is again the c jet tagging purity, and $\epsilon_{\text{tag}}$ is the tagging efficiency. After correcting for tagging efficiency and purity, the c jet \pt spectrum is obtained. This spectrum is then passed through a singular value decomposition (SVD)~\cite{SVD} unfolding procedure, as implemented by the {RooUnfold}~\cite{RooUnfold} package to remove the jet resolution effects.

\section{Systematic uncertainties and cross checks}
Systematic uncertainties for this analysis are divided into two primary categories: charm tagging and jet reconstruction.
\subsection{Tagging systematic uncertainties}
A number of systematic checks on the charm-tagged spectrum are considered, including varying the SSVHP working point,
calculating the c tagging efficiency using the JP tagger method instead of obtaining the value from simulation,
varying the gluon splitting fraction in the MC sample,
varying the MC templates within their statistical uncertainties,
and finally reweighting and varying the D meson decay parameters within the uncertainties of the world average in the simulation~\cite{H1:2004}.

The tagger working point is varied over the discriminator working point region where the use of a discriminator enhances the c jet purity.  With a very loose discriminator selection, the c jet purity is slightly enhanced relative to an unbiased sample, while a very tight selection removes the great majority of both light parton and c jets such that the b jets dominate the sample.  There is a narrow window in which the c jet purity is larger than in an unbiased sample, corresponding to the SSVHP discriminator values between 1.2 and 2.4. At its peak, the SSVHP tagger enhances the c jet purity from around 10\% to around 30\%.  To test the stability of the SSVHP tagger, multiple template fits to the corrected secondary vertex mass are performed, varying the working point of the tagger in steps of 0.2 units over this range and calculating the effective standard deviation from all working point variations.  This leads to a 2--5\% uncertainty, depending on jet \pt. An uncertainty is derived from the difference between the tagging efficiency as obtained from simulation and via fits to the JP tagger discriminator from Eq.~(\ref{eq:JPeff}). The differences in tagging efficiency between the \PYTHIA 6 estimation and using the JP tagger stem primarily from statistical fluctuation in the templates, along with a slight effect from a polynomial smoothing of these uncertainties as a function of \pt. These differences introduce a 5--15\% uncertainty, also as a function of \pt.

One of the primary theoretical unknowns in heavy-flavor jets is the impact of higher-order corrections, such as gluon splitting, and how these effects manifest themselves in these fits.  To account for this, the gluon splitting fraction in simulation is varied by 50\% up or down and the distributions of corrected secondary vertex mass are refit to the modified MC templates, where both $\Pg\to\ccbar$ and  $\Pg\to\bbbar$ splitting events are considered.  The numerical value of 50\% is used to cover observed discrepancies across various MC generators as well as discrepancies of MC generators to data, though these are primarily driven by b jet studies, where data is available.  The \PYTHIA 6 generator shows a gluon splitting contribution of about 35\%, whereas the \PYTHIA 8 generator shows a much larger contribution of around 60\%~\cite{vitev:2013aa}. Furthermore, measurements of b-dijet angular correlations in 7\TeV pp collisions show significant deviation between data and simulation as well as across generators for small dijet angular separation ($\Delta R$) values, where gluon splitting effects dominate~\cite{BPH-10-010}.  It is assumed that gluon splitting effects are as uncertain for c jets as they are for b jets. Overall, systematic uncertainty from the variation of the gluon splitting contribution is an appreciable effect in both pPb and pp collisions, though less than 15\%.

The template statistical uncertainty is accounted for by varying the distributions of light parton, c, and b jets from MC within their statistical uncertainties using a parametric MC study.
The uncertainty is estimated by fitting a Gaussian distribution to the fluctuations in purity, where the Gaussian width is used as the uncertainty value.  These values are \pt-dependent, ranging from 5\% at intermediate \pt to around 10\% at low ($\approx$60\GeVc) and high \pt ($\approx$300\GeVc).

This analysis accounts for the possibility that the \PYTHIA simulation does not accurately reproduce the D meson decay kinematics.  Since a secondary vertex that corresponds to a decay involving at least three particles is required in order to tag jets, the influence of the D meson decay parameters is studied by reweighting both the relative charm quark fragmentation and the successive D meson decay parameters in simulation to match the world average values from previous experiments.  We find that reweighting and varying these values within their uncertainties leads to a 5.5\% effect, independent of the jet \pt, collision species, and collision energy.

The contributions from each source of systematic uncertainty are summed in quadrature to obtain an overall systematic uncertainty from c jet tagging. When summed, these tagging uncertainties lead to a 10--12\% uncertainty on the fraction of charm quark jets (c jet fraction) in pp collisions, and a 10--20\% uncertainty in pPb collisions, where the majority of the extra uncertainty in pPb relative to pp comes from the JP-tagger calibration and additional unavoidable coupling of statistical fluctuations in data to the systematic uncertainty calculation at high-\pt.

\subsection{Jet reconstruction systematic uncertainties}
Additional uncertainties stem from jet reconstruction.  Jet energy corrections are derived from simulation samples and via energy balance measurements using photon+jet events. The residual non-closure of the corrections leads to a jet energy scale uncertainty ranging from 2--3\%, depending on \pt and $\eta$.  In addition, the effect of jet resolution is calculated by first smearing MC jets to match distributions of jet resolution in data, and then by using a parameterized MC study, which leads to an uncertainty of about 5\%.
The SVD unfolding procedure is cross-checked by comparing to alternative unfolding methods, including D'Agostini's method~\cite{dagostini:1994}, and by varying the raw simulated spectrum, known as the ``truth" spectrum. The uncertainty on the unfolding procedure is around 5\%, while a 4\% uncertainty is found for the simulation of the ``truth" spectrum shape. Together, all these reconstruction-based uncertainties are added in quadrature and total between 12--15\% in pPb collisions and around 15\% in pp collisions.  Finally, the integrated luminosity measurement of the pPb data has an uncertainty of 3.6\%, while the corresponding uncertainties in pp data at 2.76 and 5\TeV are 3.7 and 3.6\%, respectively. As the uncertainties from the jet energy resolution, luminosity, unfolding, and the ``truth" spectrum are canceled in the c jet fraction measurement, they are applied only to the cross section measurement.

\section{Results}

\begin{figure}[htbp]
\centering
\includegraphics[width=0.49\textwidth]{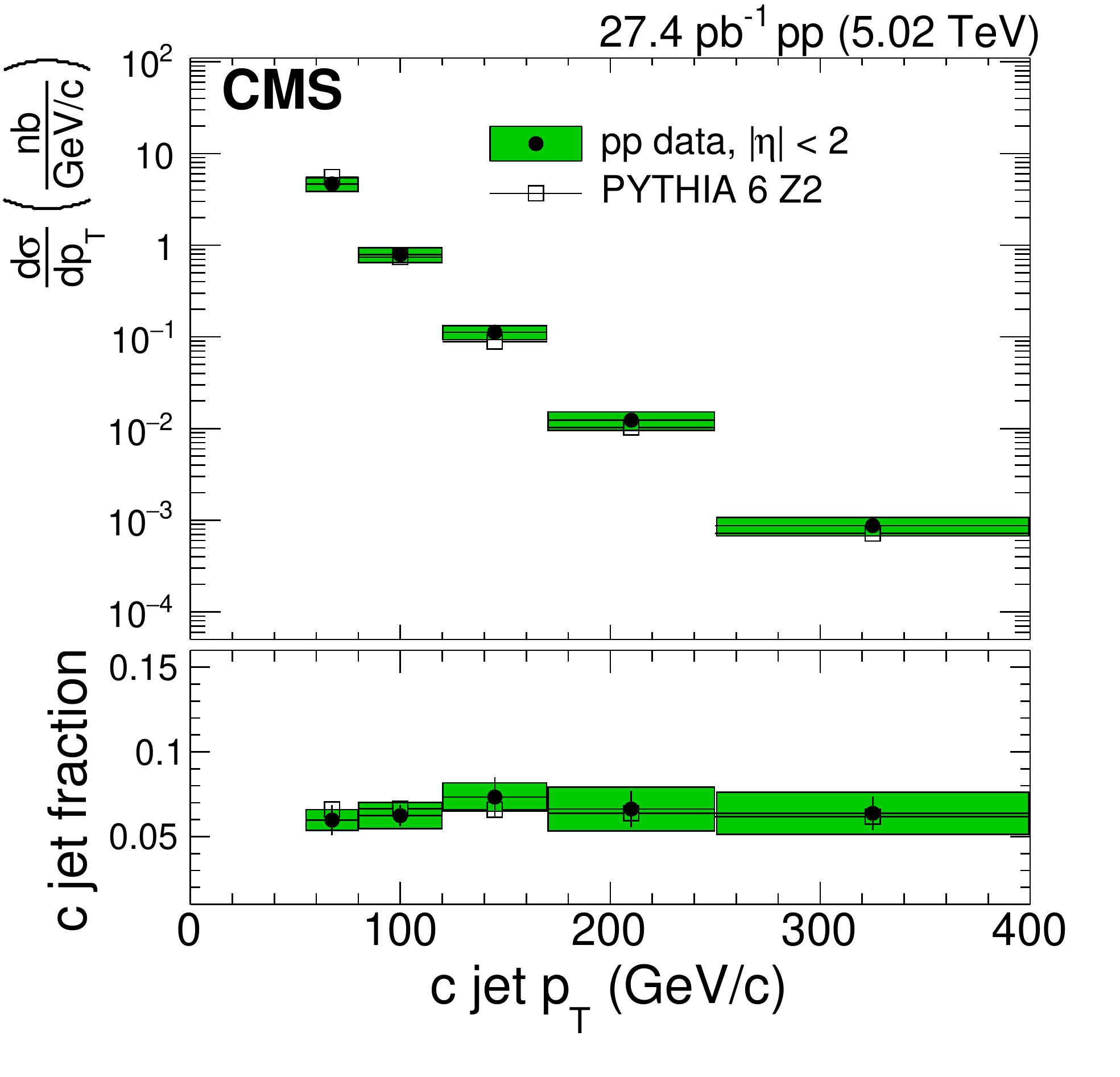}
\includegraphics[width=0.49\textwidth]{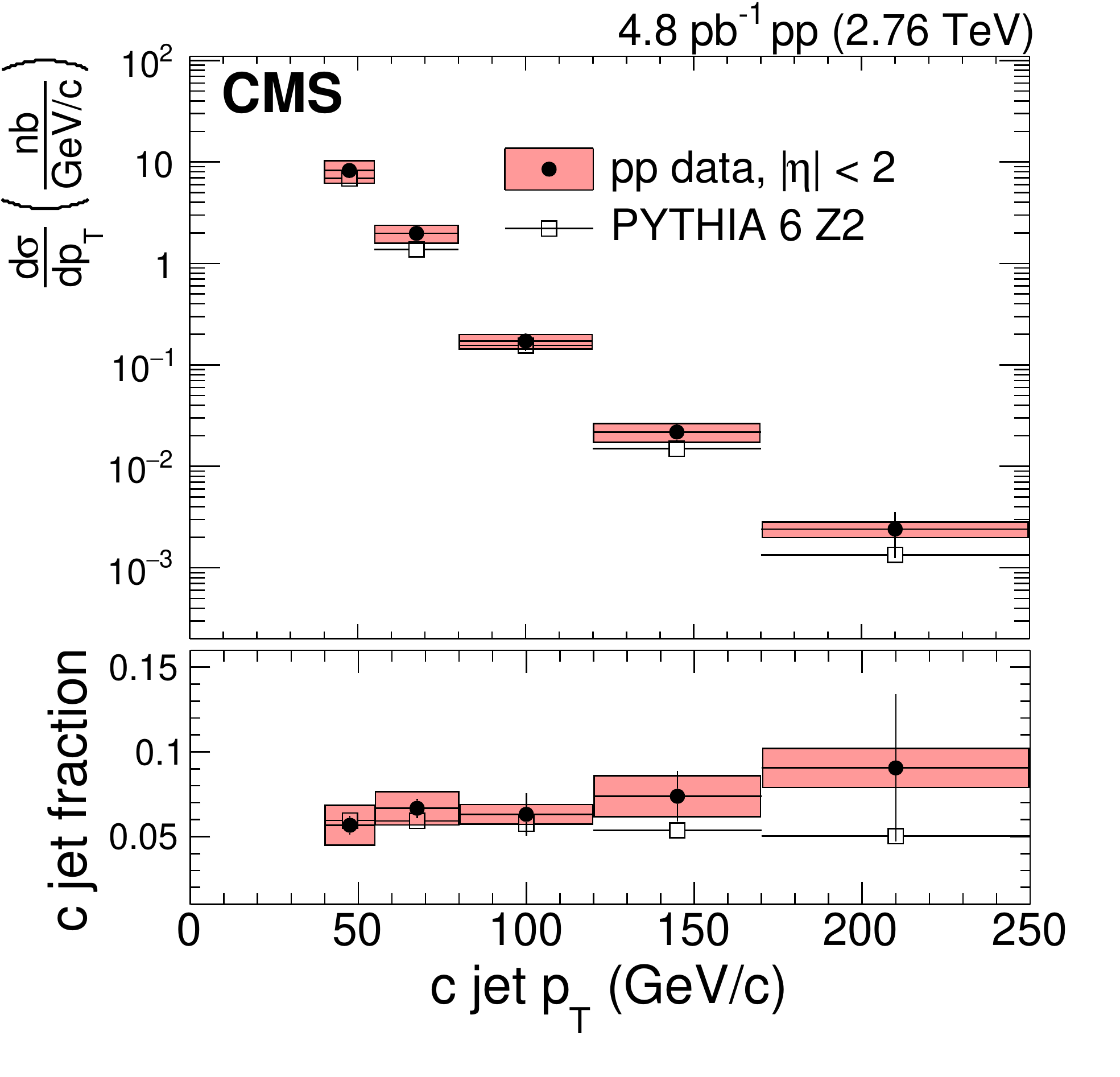}
\caption[c jet fraction in pp and pPb]{\label{fig:cFrac}
The c jet cross sections (top) and fraction (bottom) as a function of c jet \pt for 5.02\TeV (\cmsLeft) and 2.76\TeV pp data (\cmsRight), compared to predictions from \PYTHIA 6. Systematic uncertainties are shown as filled boxes.
}
\end{figure}

\begin{figure}[htbp]
\centering
\includegraphics[width=\cmsFigWidth]{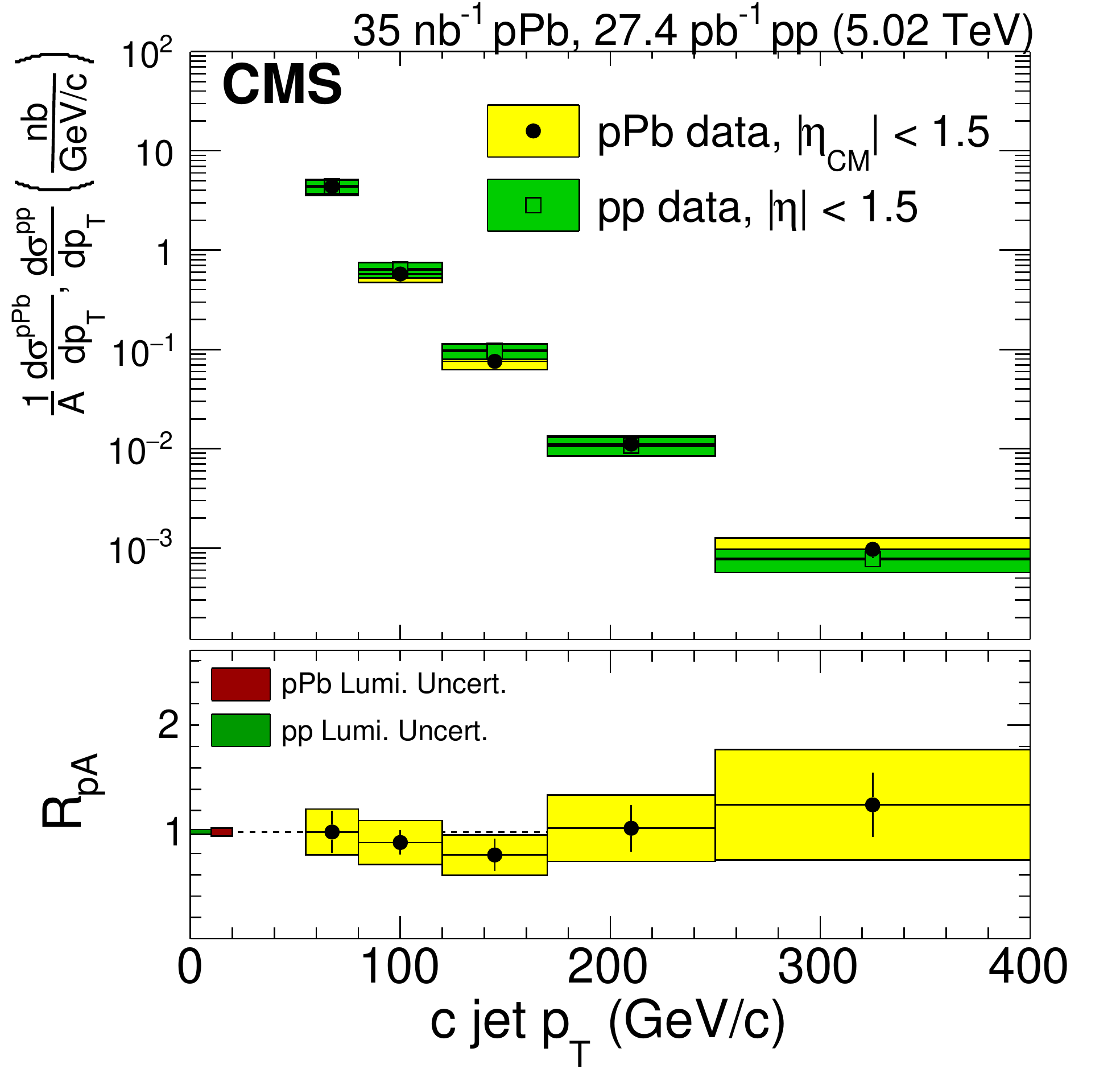}
\caption[c jet spectra in pp and pPb]{\label{fig:cSpec}
The c jet cross section (top) and \RpA~(bottom) as a function of c jet \pt for 5.02\TeV pPb and pp data.  Statistical uncertainties are solid black lines, while systematic uncertainties are shown as filled colored boxes. Integrated luminosity uncertainties for pp and pPb data are shown as filled boxes around unity.
}
\end{figure}

The c jet \pt cross section in pp collisions are shown in Fig.~\ref{fig:cFrac} for 5.02\TeV (\cmsLeft) and 2.76\TeV (\cmsRight) collisions.  The data are corrected for jet resolution by a singular value decomposition (SVD) unfolding procedure. Both cross sections are compared to predictions from the Z2 tune of \PYTHIA 6. The bottom panels of Fig.~\ref{fig:cFrac} show the c jet fraction, that is, the total number of charm jets relative to the number of inclusive jets, in pp for both collision energies. A comparison of the c jet fractions at 2.76 and 5.02\TeV suggests that the collision energy dependence of the c jet fraction is small if any and the two measurements are consistent with each other within systematic uncertainties. In addition, data from both energies confirm the \PYTHIA predictions.

The c jet cross sections as functions of \pt are shown in the top panel of Fig. \ref{fig:cSpec} for pPb and pp collisions at 5.02\TeV.
The cross sections are normalized by the total integrated luminosity of the sample.  The pPb c jet cross section is also scaled by the mass number of lead (A=208)
which normalizes the pPb measurement per binary nucleon-nucleon collision, as predicted by the Glauber model \cite{Glauber, Glauber2}. This additional scaling allows for a direct comparison of the pPb data to the pp data at the same center-of-mass energy. The direct comparison is known as the \RpA~value, which is defined as:

\begin{equation}
R_{\Pp\mathrm{A}} = \frac{1}{A}\frac{\rd\sigma^{\Pp\mathrm{Pb}}/\rd \pt}{\rd\sigma^{\Pp\Pp}/\rd\pt}.
\end{equation}

In the bottom panel of Fig.~\ref{fig:cSpec}, the c jet \RpA~value is calculated at 5.02\TeV. We observe \RpA~values consistent with unity for all \pt bins, suggesting that initial state nuclear modification effects are small for c jets at large \pt, confirming perturbative QCD predictions indicating such behavior. This absence of initial state effects is consistent with similar CMS observations for b and inclusive jets~\cite{HIN-14-007,HIN-14-001}.  Fitting a constant to the pPb c jet \RpA~\pt~distribution yields $\RpA=0.92\pm0.07\stat\pm0.11\syst$.

\section{Summary}
The transverse momentum differential cross section for c jets has been obtained for pPb collisions at $\sqrtsNN= 5.02$\TeV, as well as for pp collisions at $\sqrt{s} = 2.76$ and 5.02\TeV. The c jet fraction of $\approx$6\% is consistent with \PYTHIA simulations for pp collisions at both center-of-mass energies. By comparing the cross sections for pPb and pp collisions, a \pt-independent \RpA~value of $0.92\pm0.07\stat\pm0.11\syst$ is observed for c jets at 5.02\TeV, indicating that no significant jet energy modification is present in pPb collisions for c jets with $\pt>55$\GeVc. These measurements indicate that proton-lead initial state effects on c jets between 55--400\GeVc are small and that charm jet quenching in lead-lead collisions should not be influenced by such effects.

\begin{acknowledgments}
We congratulate our colleagues in the CERN accelerator departments for the excellent performance of the LHC and thank the technical and administrative staffs at CERN and at other CMS institutes for their contributions to the success of the CMS effort. In addition, we gratefully acknowledge the computing centers and personnel of the Worldwide LHC Computing Grid for delivering so effectively the computing infrastructure essential to our analyses. Finally, we acknowledge the enduring support for the construction and operation of the LHC and the CMS detector provided by the following funding agencies: BMWFW and FWF (Austria); FNRS and FWO (Belgium); CNPq, CAPES, FAPERJ, and FAPESP (Brazil); MES (Bulgaria); CERN; CAS, Ministry of Science and Technology of the People’s Republic of
China, and NSFC (China); COLCIENCIAS (Colombia); MSES and CSF (Croatia); RPF (Cyprus); SENESCYT (Ecuador); MoER, ERC IUT, and ERDF (Estonia); Academy of Finland, MEC, and HIP (Finland); CEA and CNRS/IN2P3 (France); BMBF, DFG, and HGF (Germany); GSRT (Greece); OTKA and NIH (Hungary); DAE and DST (India); IPM (Iran); SFI (Ireland); INFN (Italy); MSIP and NRF (Republic of Korea); LAS (Lithuania); MOE and UM (Malaysia); BUAP, CINVESTAV, CONACYT, LNS, SEP, and UASLP-FAI (Mexico); MBIE (New Zealand); PAEC (Pakistan); MSHE and NSC (Poland); FCT (Portugal); JINR (Dubna); MON, ROSATOM, RAS, RFBR and RAEP (Russia); MESTD (Serbia); SEIDI and CPAN (Spain); Swiss Funding Agencies (Switzerland); MST (Taipei); ThEPCenter, IPST, STAR, and NSTDA (Thailand); TUBITAK and TAEK (Turkey); NASU and SFFR (Ukraine); STFC (United Kingdom); DOE and NSF (USA).

\hyphenation{Rachada-pisek} Individuals have received support from the Marie-Curie program and the European Research Council and EPLANET (European Union); the Leventis Foundation; the A. P. Sloan Foundation; the Alexander von Humboldt Foundation; the Belgian Federal Science Policy Office; the Fonds pour la Formation \`a la Recherche dans l'Industrie et dans l'Agriculture (FRIA-Belgium); the Agentschap voor Innovatie door Wetenschap en Technologie (IWT-Belgium); the Ministry of Education, Youth and Sports (MEYS) of the Czech Republic; the Council of Science and Industrial Research, India; the HOMING PLUS program of the Foundation for Polish Science, cofinanced from European Union, Regional Development Fund, the Mobility Plus program of the Ministry of Science and Higher Education, the National Science Center (Poland), contracts Harmonia 2014/14/M/ST2/00428, Opus 2014/13/B/ST2/02543, 2014/15/B/ST2/03998, and 2015/19/B/ST2/02861, Sonata-bis 2012/07/E/ST2/01406; the Thalis and Aristeia programs cofinanced by EU-ESF and the Greek NSRF; the National Priorities Research Program by Qatar National Research Fund; the Programa Clar\'in-COFUND del Principado de Asturias; the Rachadapisek Sompot Fund for Postdoctoral Fellowship, Chulalongkorn University and the Chulalongkorn Academic into Its 2nd Century Project Advancement Project (Thailand); and the Welch Foundation, contract C-1845. \end{acknowledgments}
\bibliography{auto_generated}

\cleardoublepage \appendix\section{The CMS Collaboration \label{app:collab}}\begin{sloppypar}\hyphenpenalty=5000\widowpenalty=500\clubpenalty=5000\textbf{Yerevan Physics Institute,  Yerevan,  Armenia}\\*[0pt]
A.M.~Sirunyan, A.~Tumasyan
\vskip\cmsinstskip
\textbf{Institut f\"{u}r Hochenergiephysik,  Wien,  Austria}\\*[0pt]
W.~Adam, E.~Asilar, T.~Bergauer, J.~Brandstetter, E.~Brondolin, M.~Dragicevic, J.~Er\"{o}, M.~Flechl, M.~Friedl, R.~Fr\"{u}hwirth\cmsAuthorMark{1}, V.M.~Ghete, C.~Hartl, N.~H\"{o}rmann, J.~Hrubec, M.~Jeitler\cmsAuthorMark{1}, A.~K\"{o}nig, I.~Kr\"{a}tschmer, D.~Liko, T.~Matsushita, I.~Mikulec, D.~Rabady, N.~Rad, B.~Rahbaran, H.~Rohringer, J.~Schieck\cmsAuthorMark{1}, J.~Strauss, W.~Waltenberger, C.-E.~Wulz\cmsAuthorMark{1}
% ...
\vskip\cmsinstskip
% ...
1:~~Also at Vienna University of Technology, Vienna, Austria\\
% ...

\end{sloppypar}
\end{document}